# New scattering mechanism of acoustic phonons in relaxor ferroelectrics: the case of $KTa_{1-x}Nb_xO_3$

J.Toulouse[1(a)], E.Iolin[1], B.Hennion[2], D.Petitgrand[2], G.Yong[3], and R.Erwin[4]

[1]*Physics Department, Lehigh University, Bethlehem, PA, USA*
[2]*Laboratoire Leon Brillouin, CEA, Saclay, France*
[3]*Towson University, Towson, MD, USA*
[4]*National Center for Neutron Research, NIST, Gaithersburg, MD, USA*

PACS: 77.80.Jk, 78.70.Nx, 63.22.-m, 63.20.kg

ABSTRACT - The complex interaction between transverse acoustic (TA) phonon, transverse optic (TO) phonon and polar nano-domains (PND) in the relaxor ferroelectric $KTa_{1-x}Nb_xO_3$ (KTN) is studied by means of high resolution diffuse and inelastic neutron scattering. The experimental results and a comparison with lead relaxors, suggest a new scattering mechanism of the TA phonon by localized modes in PNDs. A theoretical model is developed, which accurately predicts the evolution of the TA damping with temperature and wavevector. Such a mechanism suggests the possible use of high frequency acoustic modes for the study of nanocomposite materials.

Relaxor ferroelectrics represent a now well recognized subgroup of highly polarizable compounds with substitutional disorder in the form of off-center ions that introduce random electric dipole moments throughout the lattice. With decreasing temperature, correlations develop between these, leading to the observation of polar nanoregions (PNR) below the so called Burns temperature, $T \approx T_d$. At yet a lower temperature, $T^* < T_d$, the dielectric constant begins to exhibit a characteristic frequency dispersion or relaxor behavior. This suggests a distinction between polar nanoregions, resulting from dynamical polar correlations, and polar nanodomains (PND) when these correlations become static or long lived resulting in local distortions and elastic diffuse scattering as shown below.[1,2] An order of magnitude for the minimum life time of these regions can be obtained from the dielectric constant measurements shown later in the paper. The formation of PNDs of finite size ***a*** with a different structure than the surrounding lattice can be viewed as a local ferroelectric phase transition or the condensation of the soft optic mode in a finite interval of ***q~1/a***. [3]. This finite size structural change is accompanied by the appearance of elastic diffuse scattering which provides an estimate of the size of PNDs, 20Å-50Å depending on the system and the temperature. The relaxor behavior is due to the orientational relaxation of these PNDs. The other essential contribution to the dynamics of relaxors comes from their lattice dynamics, the soft transverse optic (TO) phonon and the low frequency transverse acoustic (TA) phonon. In addition, new local modes intrinsic to the PNDs may also be expected to appear, with which the TO and TA phonon should interact and through which they should be observable. In the present Letter, we report high resolution inelastic neutron scattering results of the TA phonon in $KTa_{1-x}Nb_xO_3$ (KTN) which provide the first experimental evidence that such local modes indeed exist in relaxors.

Among known relaxors, KTN is a particularly useful model system because of both its chemical and structural simplicity (Ta and Nb both with valence 5+ and approximately same radius), in particular the absence of short range chemical ordering when compared with lead relaxors such as $PbMg_{1/3}Nb_{2/3}O_3$ (PMN) and $PbZn_{1/3}Nb_{2/3}O_3$ (PZN)[4] The TO and TA phonons and diffuse scattering have been extensively studied in PMN, PZN, PMN-PT and PZN-PT (with added $PbTiO_3$) and shown to interact between themselves and with the PNDs independently [5,6]. The authors of Ref.6 explained the observed TA phonon damping by an interaction with PNR and reported strong damping when the TA phonon was excited near a point (h, k, l) in the Brillouin zone (BZ) where diffuse scattering was also strong. The results presented below show that the TA phonon can also be scattered by PNDs near a point where diffuse scattering is weak, albeit through a new different scattering mechanism based on a three-way interaction TA - TO phonon and local modes in PNDs. A theoretical model is also developed that accurately describes these results.

In our study, the presence of PNDs in a $KTa_{0.85}Nb_{0.15}O_3$ (KTN15) crystal was first evidenced through elastic diffuse scattering, which directly reveals the development of local structural order with a lower symmetry than that of the surrounding lattice. These measurements were carried out on the BT2 spectrometer at the neutron facility at NIST with collimations of 60'-40'-40'-80' and neutron energy of 14.7meV. Pyrolytic graphite was used to filter out harmonics. The scattering plane was the (100)-(011) plane and elastic scattering was measured upon warming around the (110) Bragg peak in transverse <001> scans (in the (100)-(110) scattering plane). In fig.1, diffuse scattering is clearly visible below 160K, indicating the formation of PNDs. The Bragg peak maximum increases rapidly below 140K due to the relief of primary and secondary extinction caused by atomic plane distortions; the diffuse scattering intensity goes through a clear maximum around 130K. Simultaneously, the full width at half maximum of the diffuse scattering goes through a minimum (not shown), from which we extract the maximum correlation length, ξ(T=130K)≈10 unit cells ≈39.5Å. As in PMN and PMN-PT, neutron diffuse scattering in KTN occurs at (1,1,0) *but not* around (2,0,0).[7]

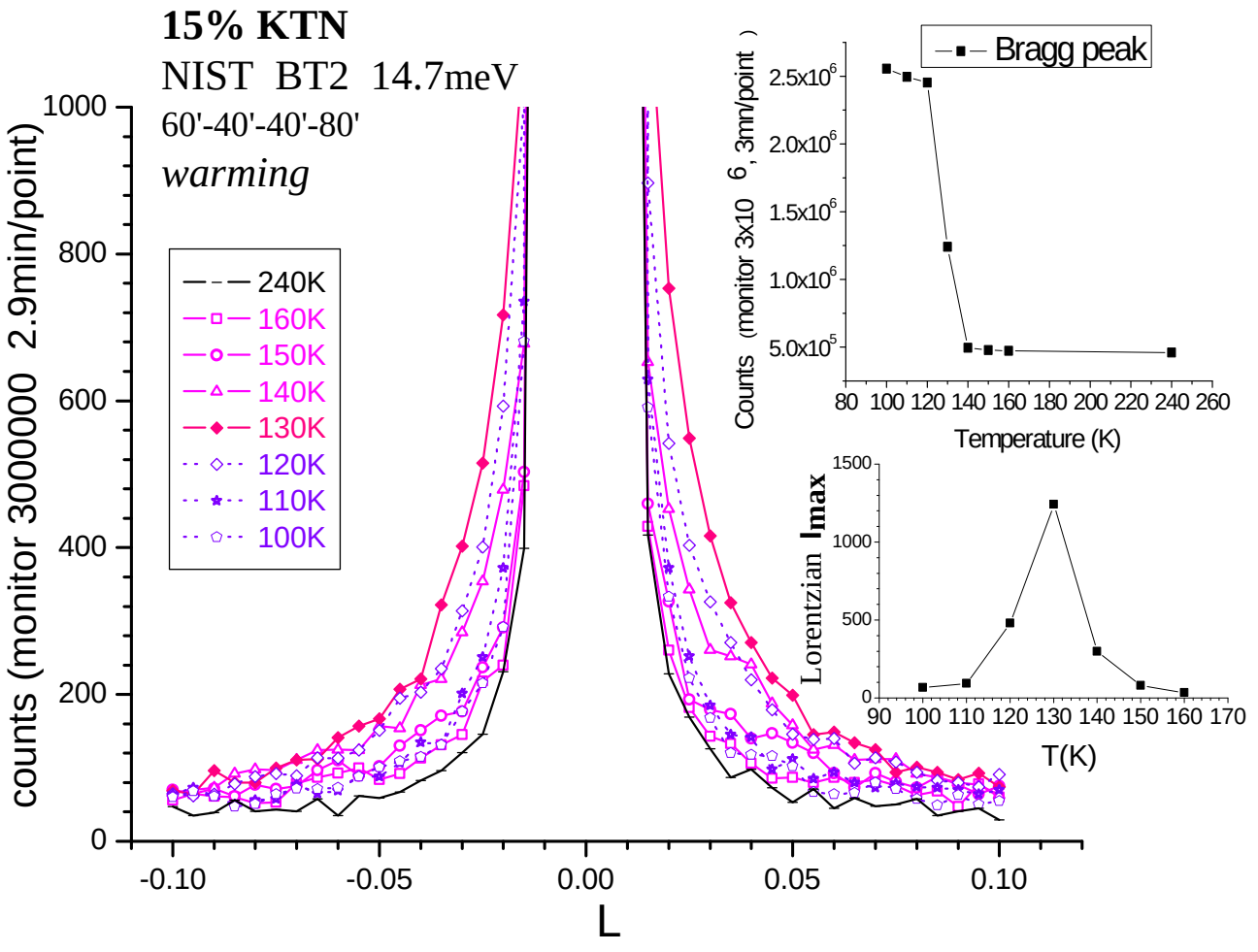

***Fig.1 (110) Elastic diffuse scattering for different temperatures; inset: Bragg and diffuse peak intensities vs T.***

Measurements of the transverse acoustic (TA) phonon were made on the SP4F2 cold neutron triple axis spectrometer at the Laboratoire Leon Brillouin in Saclay (France), near (200) in the (100)-(010) scattering plane with the phonon propagating in the (010) direction. The effective collimations used were, horizontally: 273'-27'-40'-40', vertically: 51'-69'-137'-275' and the final neutron wavevector was $1.64Å^{-1}$ from q=0.025 to q=0.09, providing a high resolution of ~0.2 meV (0.05THz), and $2.662Å^{-1}$ at q=0.12 and 0.16, providing a resolution of 1meV. Pyrolytic graphite was also used to filter the harmonics. Several spectra are shown in Fig.2. The fitted curves were obtained using a damped harmonic oscillator description for the phonon, including thermal population and convoluted with the spectrometer resolution. This convolution is very important in KTN, and possibly in other perovskite relaxors, because of the large anisotropy of the phonon dispersion surfaces, with deep valleys along <100> directions and high ridges on either side. For a given ***q***, the same functional form was used in the convolution at different temperatures.

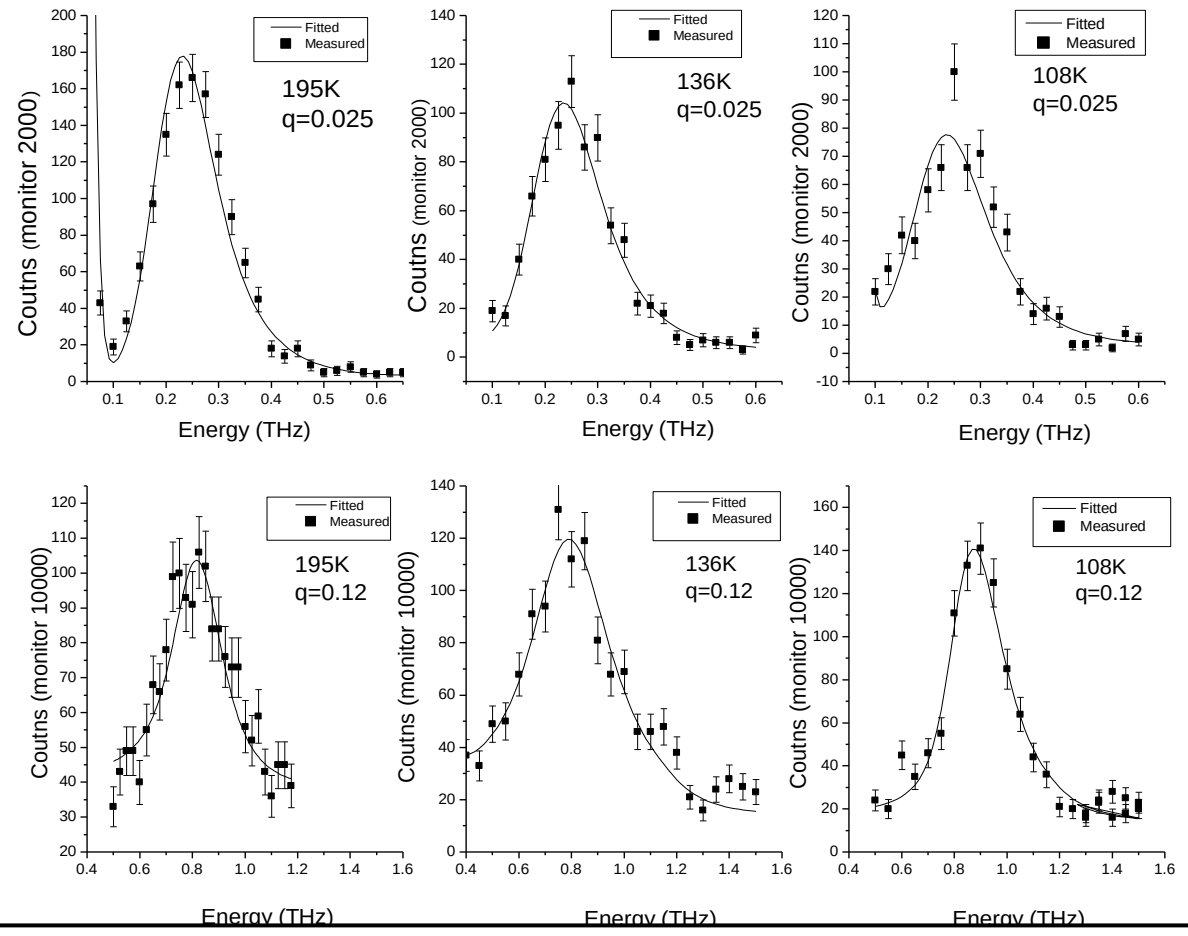

***Fig.2 Transverse acoustic phonon spectra (experimental data and fitting curves) for q=0.025and q=0.12 at different temperatures, determined relative to the (200) Bragg reflection that in the region where diffuse scattering is very small.***

The temperature dependence of the phonon damping is shown in fig.3. For small wave vectors, the phonon width is relatively flat and independent of temperature. For $\boldsymbol{q}\geq0.07$, the phonon width shows a clear maximum around 136K, a temperature at which the PNDs have appeared (see diffuse scattering in Fig.1). Most significant, however, is the fact that the phonon width or damping decreases significantly at lower temperature. This temperature profile of the TA damping indicates that the lattice is relatively ordered at high temperatures, becomes partially disordered at intermediate temperatures, when the PNRs and then the PNDs are present, and finally returns to a more homogeneous although still distorted state (large Bragg intensity) at lower temperature when the PNDs grow and merge. Although not shown here, the TO mode measured near the same (200) reflection all but disappears from the spectrum in the temperature region of the maximum TA damping, a phenomenon also observed in lead relaxors and designated as "waterfall". Koo et al. have reported very similar results in PMN-20PT [8]. By contrast, in an almost pure $KTaO_3$ (with 1% lithium), the phonon width was found to decrease continuously between 250K and 50K and no elastic diffuse scattering was observed. [9] These results do seem to suggest that the TA broadening and damping maximum observed in KTN15 and PMN-20PT are linked to the presence of PNRs or PNDs. (see discussion below). The dielectric constant curves shown in the inset confirms the relaxor behavior and indicates the absence of a clear phase transition. The frequencies of the curves displayed also give an order of magnitude for the minimum lifetime of the PNDs.

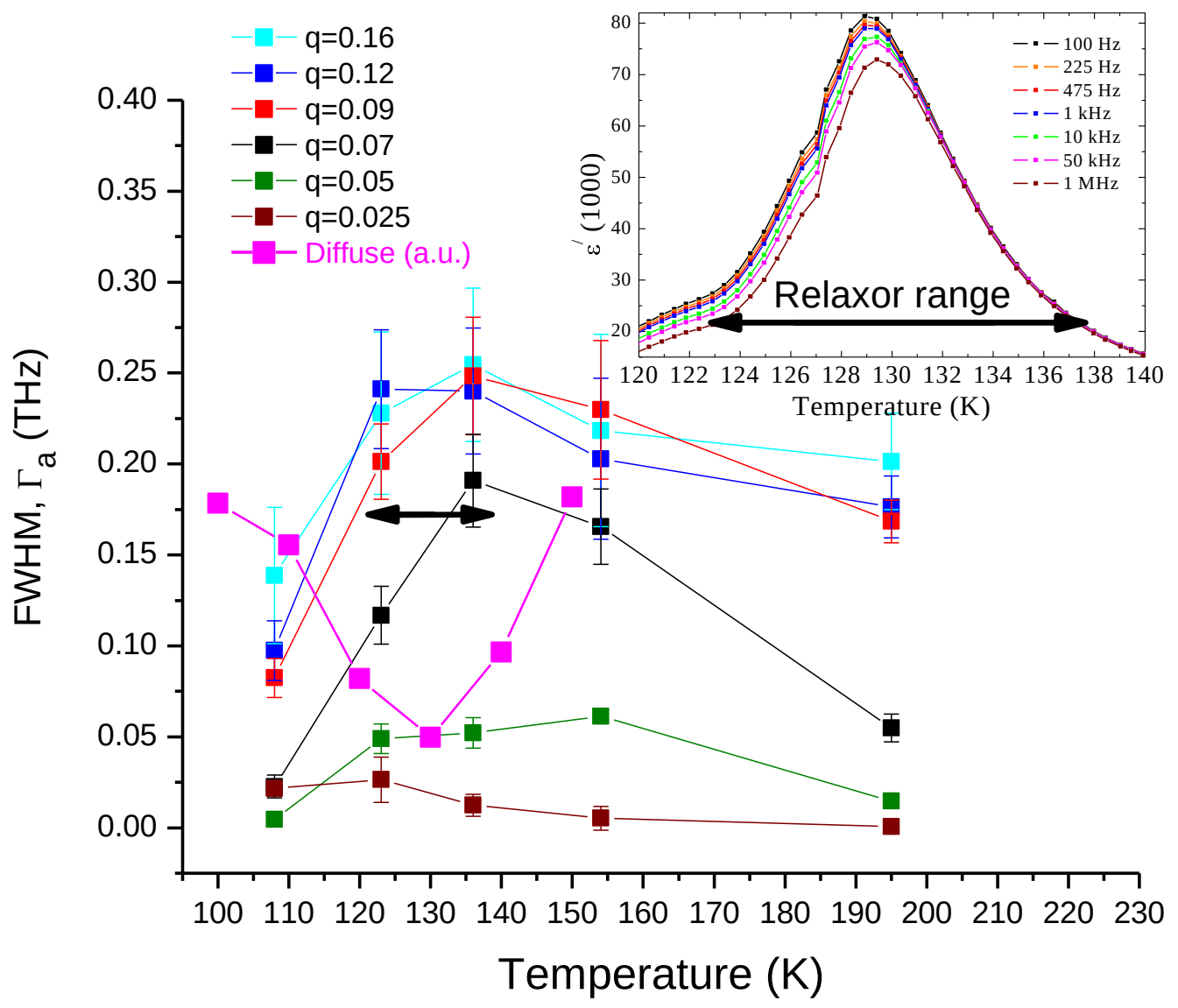

***Fig.3 Damping (FWHM) of the transverse acoustic (TA) mode vs temperature for different q values; inset: dielectric constant vs. T***

In Fig.4 we present the $\boldsymbol{q}$ dependence of the damping, which also reveals an interesting trend. At all temperatures, the phonon width exhibits a step increase around $\boldsymbol{q}$=0.07, goes through a shallow maximum at $\boldsymbol{q}$=0.09-0.12 and remains high beyond. The step increase is moderate at 195K, the highest temperature measured, maximum at intermediate temperatures around 140K and significantly smaller at the lowest temperature measured of 108K. Such a step increase with maximum around $\boldsymbol{q}$=0.09-0.12 suggests scattering of the TA phonon by homogeneities on a scale of approximately 8÷11 unit cells, which is very similar to the coherence length of ~10 unit cells obtained earlier from elastic diffuse scattering. In Fig.4b, we take a closer look at the $\boldsymbol{q}$ dependence of the damping on a log-log scale and note two distinct slopes. For small $\boldsymbol{q}$, $\boldsymbol{\Gamma_a} \propto \boldsymbol{q}^4$ at the highest and lowest temperatures, 195K and 108K, and $\boldsymbol{\Gamma_a} \propto \boldsymbol{q}^2$ at intermediate temperatures, with a cross-over around T~154K. What is particularly significant in these results is the fact that the change in power law occurs precisely near the temperature where elastic diffuse scattering reveals the formation of PNDs. In addition, the cross-over wavevector corresponds to scattering objects with size $\boldsymbol{a}$~8÷11 lattice units, or 32÷44 Å, quite comparable with the size inferred from the elastic diffuse scattering measurements.

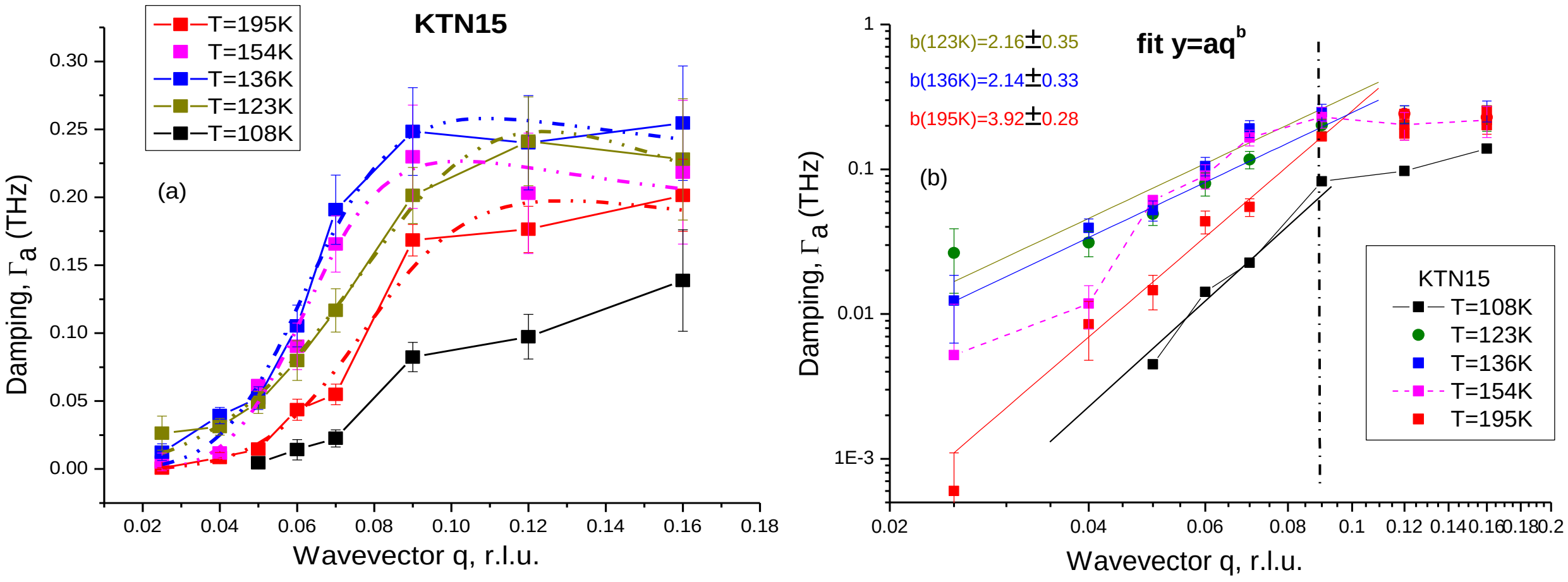

***Fig.4 Damping $\Gamma_a$ vs wavevector q on a) a linear scale with fit (dash line) using Eq.(4) in the text; b) on a log-log scale with linear fit to extract the exponent n, $q^n$. 154K clearly appears as a cross-over temperature.***

The central experimental fact from our results is that, despite little or no diffuse scattering near the (200) reflection in KTN [7], the TA phonon is still seen to broaden with decreasing temperature, to exhibit maximum damping in the temperature range in which the PNDs form and to narrow again at low temperature. Similar results obtained in PMN-20PT [8] were already mentioned above. These results should also be compared with those of Stock et al. [6] who measured the TA phonon in PMN near (200) and (220) as well as near (110). Given their experimental resolution, the TA phonon did not appear to broaden significantly near the first two reflections, where little or no diffuse scattering is observed, but did broaden considerably near the (110) reflection, where strong elastic diffuse scattering is present. These authors therefore concluded to a separate coupling between the acoustic mode and the diffuse scattering near (110) and to the usual TA-TO coupling near (220) and (200). This conclusion is surprising because, as seen from their scattering geometry in the reciprocal plane, the TA phonon excited near (220) and (110) *should be one and the same TA phonon* and should therefore equally interact with the PNDs near both reflections, irrespective of the strength of the diffuse scattering at the particular point of the reciprocal lattice. One must conclude that *the two TA phonons excited respectively near the two reflections are in fact not the same*, one being excited outside PNDs as near (200) and the other inside PNDs as near (110). This conclusion is the starting point of our analysis. It is important to note that no TA damping maximum was observed in an almost pure KTaO3 crystal (1% Li) that also did not show any diffuse scattering [9]. Therefore the TA damping maximum observed in the present KTN15 results cannot be simply due to the usual TA-TO coupling.

In an almost perfect crystal, elastic diffuse scattering arises from small deviations from the perfect structure surrounding impurity centers or small aggregates. The case of KTN15 or PMN-20PT may however be quite different, as the volume of PNDs is not small and can represent a significant fraction of the whole volume, possibly as much as 30%.[10] Elastic diffuse scattering in relaxors may therefore arise from Bragg scattering within finite size PNDs (20Å-50 Å) that have a different structure than the outside lattice. And the absence of diffuse scattering around (200) is evidence that the static structure factor near this reciprocal lattice point is the same inside and outside PND. As explained below, this implies that TA phonon will be excite near (200) in the same way as in the perfect crystal without PNDs. However, the PNDs clearly have an effect on the TA damping near (200) (not seen in an almost pure crystal without diffuse scattering [9]), which must be of a different nature than their effect on the TA damping near other reflections such as (110) where diffuse scattering is observed. Therefore, the TA phonon can interact differently with PNDs at different reciprocal lattice points. Indeed, the dynamic structure factor that determines the intensity scattered by long wavelenth TA phonons also contains the static structure factor: $\boldsymbol{F_{pc}(G,\mathbf{q}) = F(G)(G \bullet \mathbf{u})}$ with $\boldsymbol{F(G) = \sum_{\alpha} b_{\alpha} \exp(iG \bullet \mathbf{r}_{\alpha})}$, where $\boldsymbol{u}$ is the displacement, *F(G)* is the static structure factor and $\boldsymbol{b_\alpha}$ and $\boldsymbol{r_\alpha}$ are the scattering length and position of the $\boldsymbol{\alpha}^{th}$ nucleus in the unit cell. The coexistence of the large damping of the TA phonon and strong diffuse scattering reported by Stock et al.[6] near the (110) reflection in PMN indicates that this TA phonon is excited inside the PNDs. Conversely, the absence of diffuse scattering near (200) or (220) would indicate that, due to a small static structure factor, $\boldsymbol{F(G)}$, the TA phonon is not directly excited inside PNDs. As shown by the data, the TA broadening and damping maximum is nevertheless related to the presence of the PNDs, and must therefore be scattered by them through an indirect scattering mechanism. Early on Axe et al [11] evidenced the importance of the interaction between the transverse optic and the transverse acoustic mode (TO-TA) in the [1, 0, 0] direction in the incipient ferroelectric $KTaO_3$, and this interaction is also likely to play a role in KTN15. However, a major difference with KTN is that the TA and TO branch do not cross and PNRs and PNDs are absent in $KTaO_3$ (no diffuse scattering even near (110) [9]). In KTN15 on the other hand, PNDs resulting from the condensation of the TO mode are present and localized modes with small dispersion are expected to exist.[12] The very strong damping of the TO phonon or "waterfall" phenomenon observed near (200) in KTN15 as well as in lead relaxors reveals the strong interaction between TO phonons outside and these localized TO modes inside PNDs. Similarly, the significant broadening of the TA phonon observed in lead relaxors in the presence of large diffuse scattering near (110) [13] suggests a strong interaction between the TA and localized modes inside PNDs. Such a strong interaction is not expected however for TA phonons measured near the (200) reflection since, as was explained earlier, these are not being excited inside PNDs (no diffuse scattering near (200)). Instead, a three-way interaction between the TA and TO phonons and localized TO modes in PNDs is very likely. Such a three-way interaction is most natural, deriving from the TA-TO interaction already present in pure $KTaO_3$ but incorporating the strong interaction between the TO phonons outside in the undistorted lattice and those localized inside PNDs. This indirect interaction of a "TO-dressed" TA phonon with localized TO modes would also explain the milder broadening of the TA phonon observed near (200) as compared to the strong one near (110). It is important to note, that the damping behavior of the TA phonon observed in KTN cannot be explained by a simple interaction with a soft TO mode. Indeed, as shown in Fig.11 of Ref. [14], the TO mode soften only slightly and goes through a shallow minimum in the range of temperatures where the PNDs appear. The TO phonon frequency cannot in fact be precisely determined in the region of the shallow minimum because the TO phonon peak disappears from the spectrum, the same waterfall phenomenon as observed in PMN.[5]

Today our precise knowledge of the characteristics of PNDs is too limited for a detailed calculation taking into account their shape and boundaries. Therefore, we choose to develop a quantitative model assuming an effective homogeneous medium and show that it can accurately predict the wave vector dependence determined experimentally. We assume a Landau temperature dependence of the soft optic mode gap, $\omega_0^2 = a_0(T-T_c)$ with $\boldsymbol{T_c}$ the ferroelectric transition temperature and $\boldsymbol{a_0}$ a constant parameter. The acoustic mode dispersion is, $\Omega_a = c_a q$, and the optic mode dispersion, $\Omega_o{}^2 - (\omega_o^2 + c_o^2 q^2) - i\Omega_0\Gamma_0 = 0$, in which $\boldsymbol{c_a}$ and $\boldsymbol{c_o}$ are the velocities of the acoustic and optic mode respectively and $\boldsymbol{\Gamma_0}$ is the TO damping, or more exactly the coherence decay rate of the TO modes which also includes their inhomogeneous broadening. The transverse acoustic and optic strains are written as

$$\xi_{ik} \equiv 1/2(\partial\xi_k/\partial r_i + \partial\xi_i/\partial r_k),$$
$$u_{ik} \equiv 1/2(\partial u_k/\partial r_i + \partial u_i/\partial r_k) \quad (1)$$

with $u_k$ and $\xi_k$ the *k* components of the acoustic and optic mode displacement eigenvectors respectively. The interaction Hamiltonian, $\boldsymbol{H_{int}}$, between TA and TO contains two terms. The first, $\boldsymbol{H_p}$, is similar to that considered early by Axe et al [11], and leads to the "dressed" TA phonon and is important in both paraelectric and ferroelectric phases. The second term, $\boldsymbol{H_f}$, represents the modulation by the TA strain of the ferroelectric interaction between soft mode displacements $\boldsymbol{\xi}$ or between soft mode displacement and polarization [15].

$$H_{int} = H_p + H_f, \quad H_p = 2g u_{ik}\xi_{ik}$$
$$H_f = H_{fr} + H_{fnr} \quad H_{fr} = 2f(P\xi)_{ik}u_{ik} \quad H_{fnr} = f\xi_i\xi_k u_{ik} \qquad (2)$$
$$(P\xi)_{ik} \equiv 1/2(\xi_i P_k + \xi_k P_i)$$

in which the parameters $\boldsymbol{g}$ and $\boldsymbol{f}$ are constants of interaction. The resonance term, $\boldsymbol{H_{fr}}$ is bilinear in the TA strain and TO wave displacement and proportional to the static polarization vector $\boldsymbol{P}$, and is only effective in the PND phase; the non-resonant term $\boldsymbol{H_{fnr}}$ leads (in general) to the scattering of the TA phonon by the soft mode. As shown later however, fitting of the data reveals that the soft-mode phase velocity is smaller than that of the acoustic mode, $c_o < c_a$, so that the requirement of energy and momentum conservation cannot be satisfied in a process in which an optical phonon is scattered with the emission or absorption of a single TA phonon. On the other hand, the probability of emission (absorption) of two TO phonons by the TA phonon will be small because this process would only occur in a small part of phase space. Therefore we can omit the term $H_{fnr}$.

Assuming plane waves with polarization along the *x* axis and propagating in the *y-z* plane with momentum $q$ and frequency $\omega$, the coupled equations of motion obtained can be written as:

$$(\omega^2 - c_a^2 q^2)u + [-gq^2 + if(q_y P_y + q_z P_z)]\xi = 0,$$

$$[-gq^2 - if(q_y P_y + q_z P_z)]u + (\omega^2 - c_o^2 q^2 - \omega_o^2 + i\omega\Gamma_0)\xi = 0 \qquad (3)$$

Solving these coupled equations for $q$ using perturbation theory to the first non-vanishing order with respect to $g$ and $f$, we obtain the following expression for the acoustic damping:

$$1/\tau_r = \mathrm{Im}(\delta q) \approx \frac{1}{2c_a} \frac{g^2 q_0^{\,4} + f^2 (q_{0y} P_y + q_{0z} P_z)^2}{(\omega^2 - c_o^2 q_0^{\,2} - \omega_o^2)^2 + (\omega\Gamma)^2} \Gamma_0 \equiv \frac{p_1 (q^4 + p_4 q^2)}{(q^2 - p_2^2)^2 + p_2^4 p_3^2} \qquad (4)$$

where $q=q_0+\delta q$, $q_0=\omega/c_a$ and with the last expression parameterized in terms of the fitting coefficients $p_1$-$p_4$. Eqs. (2)-(4) are written for the case of an isotropic medium but are also applicable here when the TA phonon is propagating along the high symmetry [010] or *y* direction of the cubic crystal. The first term in Eq (2), proportional to *g*, accounts for the more conventional TA-TO coupling (as per Axe et al [11]) and predicts a $q^4$ dependence of the TA phonon damping , while the second one, proportional to *f*, incorporates the specific contribution of the PNDs and predicts a $q^2$ dependence. The latter is essential when both transverse TA phonon and soft TO mode are propagating in a direction parallel to the polarization vector $P$. It is also important to note that the acoustic phonon damping is proportional to the TO coherence decay rate, $\Gamma_0$, which is in the end responsible for the acoustic damping. For small $q$ and high temperatures, the polarization fluctuations are negligible or absent, **P=0**, and $\mathrm{Im}(\delta q) \sim g^2 q^4$, as observed at 194K in Fig.4b. However, for small $q$ and low temperatures, $q_0^4 \ll q_0^2$, $P \neq 0$ and $\mathrm{Im}(\delta q) \sim f^2 q^2$, as observed below 154K. For large $q$ we can expect the acoustic damping to be independent of $q$. Therefore, this model accurately predicts the observed $q$ dependence of the TA phonon damping over the whole temperature range studied.

The damping curves presented in Fig.4a have been fitted to the parameterized form of eq. (4). From the fitting, $p_4 \equiv f^2 q^2$ is found to be very small or negligible down to T=136K. However, it makes an essential contribution below 136K, as should be expected from the local polarization, $P$, present in the PNDs at lower temperatures. We also find $p_2^2 > 0$ at all temperatures, indicating a resonant rather than a relaxation type of interaction [see $(q^2 - p_2^2)^2$ term in the denominator of eq. (4). In the expression for $p_2$, $(c_a^2 - c_o^2)\omega_0^2 = (c_a^2 - c_o^2) p_2^2 + c_a^2 \Gamma^2$, all terms on the right hand side are positive, suggesting that, on the left hand side, either $c_0 < c_a$ or $\omega_0^2 < 0$. The latter is not an option since it would indicate that the system has already condensed to the ferroelectric phase at the highest temperature. Consequently, $p_2^2 > 0$ must signify $c_0 < c_a$, or equivalently a small dispersion of the TO mode as would be expected for localized modes. This is very different from pure $KTaO_3$ in which $c_0 / c_a \sim 2$ or 3 [11, 3] and further confirms that the TO dynamics inside and outside PNDs are different. In fact, given the very large damping of the TO phonon, it might actually be more exact to regard it as a "hybrid" TO phonon, partly localized and partly delocalized and with a small dispersion.

Although it describes a different case than ours, it is worth mentioning the model used by Stock et al.[6] to fit their low energy neutron spectrum of PMN near the (110) reflection. This model follows the formalism proposed by Michel and Naudts [16] to describe the coupling between acoustic mode and molecular rotations in molecular crystals, and predicts a relaxation contribution to the TA phonon damping of the form, $1/\tau_d \sim \omega^2 \lambda / (\omega^2 + \lambda^2)$, where $\lambda$ is the relaxation frequency. This model is obviously not applicable near the (200) reflection where diffuse scattering is absent. In addition, it does not take into account a likely contribution from the TO phonon, which is explicitly taken into account in our model. Finally, Naberezhnov et al [17] also analyzed the damping of the TA phonon in PMN at 650 K in the frame of a two coupled oscillator model. Their analysis suggested the existence of a low frequency so called quasi-optic (Q-O) excitation with dispersion $\omega^2 = \omega_0^2 + Dq^2$ with $\omega_0$=0.287 THz and D=88.5 $THz^2$. Vakhrushev and Shapiro [18] also suggested the existence a second low energy TO mode in PMN, appearing in their neutron spectrum between the TA phonon and the quasi-elastic or inelastic diffuse scattering. It is possible that our weakly dispersive TO mode, which we presume is partly localized inside PNDs, might be this Q-O mode proposed in Ref. [18].

In the large $\boldsymbol{q}$ limit, large defects or inclusions have been shown to result in $\boldsymbol{q}$-independent scattering of the acoustic phonons [19]. In the present case, this suggests that the PNDs grow and eventually merge, resulting in a microstructure more uniform average lattice environment, as also indicated by the reduced acoustic phonon damping at low temperature. It is also worth noting in Fig.4b, that the TA phonon damping follows again a $\sim q^4$ law at the lowest temperature measured of 108K. This $q^4$ dependence is consistent with the fact that the low temperature phase in a mixed crystal such as KTN15 is not homogeneous (see large Bragg width) and that the TA phonon is likely to be Rayleigh scattered by the small inhomogeneities present in the low temperature phase just as it was at high temperature.

In summary, we have identified a new scattering mechanism of the transverse acoustic (TA) phonon in the relaxor ferroelectric KTN15. A high resolution neutron scattering study of the transverse acoustic (TA) phonon reveals a damping maximum in the temperature region where polar nanodomains (PND) form and a step increase in damping at a wavevector $\boldsymbol{q}$ that corresponds to the size of the PNDs determined independently from elastic diffuse scattering on the same crystal. These results are unexpected because, although the TA behavior is shown to be related to the presence of PNDs, the damping is measured near the (200) point of the reciprocal lattice, where little or no diffuse scattering is observed. These results suggest that the TA phonon is indirectly scattered by PNDs in a three-way interaction between the transverse optic (TO) and transverse acoustic (TA) phonons, or “TO-dressed” TA phonon, and localized modes intrinsic to the PNDs. We have developed a formal theoretical model which describes this three-way interaction and accurately predicts the observed wavevector dependence of the TA damping, $\sim \boldsymbol{q}^4$ at the highest and lowest temperatures, corresponding to Rayleigh scattering by small size inhomogeneities, and $\sim \boldsymbol{q}^2$ at intermediate temperatures corresponding to resonant scattering by localized modes in PNDs. The KTN15 results are also compared with similar results obtained in PMN and PMN-PT, which can also be interpreted in the same way. Finally, this work demonstrates that the THz TA phonons excited by neutrons can potentially be very effective probes of nano- or meso-scopic order in solids

ACKNOWLEDGMENTS
This work was supported for a large part (JT and EI) by the US Department of Energy under grant DE-FG-06ER46318. We thank L.A. Boatner of Oak Ridge National Laboratory for a high quality KTN crystal.